\documentclass[aps,showpacs,preprintnumbers,amsmath,amssymb]{revtex4}

\usepackage{float}
\usepackage{graphics,epsfig}
\usepackage{graphicx}
\usepackage{dcolumn}
\usepackage{bm}

\begin{document}
\baselineskip=0.8 cm

\title{ dynamical evolution of scalar perturbation in Ho\v{r}ava-Lifshitz black-hole spacetimes }
\author{Chikun Ding }
\author{Songbai chen}
\author{Jiliang {Jing}\footnote{Corresponding author, Electronic address:
jljing@hunnu.edu.cn}} \affiliation{ Institute of Physics and
Department of Physics, Hunan Normal University, Changsha, Hunan
410081, People's Republic of China \\ and \\ Key Laboratory of Low
Dimensional Quantum Structures and Quantum Control (Hunan Normal
University), Ministry of Education, P. R. China.}

\vspace*{0.2cm}
\begin{abstract}
\baselineskip=0.6 cm
\begin{center}
{\bf Abstract}
\end{center}

We study the dynamical evolution of a massless scalar perturbation
in the Ho\v{r}ava-Lifshitz black-hole spacetimes with the coupling
constants $\lambda=\frac{1}{3}$, $\lambda=\frac{1}{2}$ and
$\lambda=3$, respectively. Our calculation shows that, for the three
cases, the scalar perturbations decay without any oscillation in
which the decay rate imprints the parameter of the
Ho\v{r}ava-Lifshitz black hole. The results are quite different from
those in the Schwarzschild AdS black hole and can help us understand
more about the Ho\v{r}ava-Lifshitz gravity.

\vspace*{0.2cm}


\end{abstract}

\maketitle
\newpage
\section{introduction}

Einstein's general relativity is increasingly important in the
modern physics,  especially in the frontiers of very large distance
scales including astrophysics and cosmology. However, it is proved
to be non-renormalizable by quantum field theories. Therefore,
numerous attempts of its modification have appeared in the
literatures. Recently, Ho\v{r}ava \cite{ho1} proposes a power
counting renormalizable gravity theory in four dimensions, which
admits the Lifshitz scale-invariance in time and space rather than
Lorentz invariant theory of gravity at ultraviolet level. In the
infrared limit, the higher derivative terms do not contribute and
Ho\v{r}ava-Lifshitz gravity  reduces to standard general relativity.
Thus, Ho\v{r}ava-Lifshitz gravity can be regarded as an ultra-violet
complete theory of general relativity.

Due to these novel features, Ho\v{r}ava-Lifshitz gravity theory has
been intensively investigated
\cite{ho2,ho3,VW,klu,Nik,Nas,Iza,Vol,CH,CHZ} and its cosmological
applications have been  studied \cite{cal,TS,muk,Bra,pia,gao,KK}.
Some static spherically symmetric black hole solutions with
non-vanishing cosmological constant have been found in
Ho\v{r}ava-Lifshitz theory \cite{CY,LMP,CCO,CLS,Gho} and the
associated thermodynamic properties with those black hole solutions
have been investigated \cite{MK,Nis,CCO1,Myung}. The potentially
observable properties of black holes in the deformed
Ho\v{r}ava-Lifshitz gravity were considered by using the
gravitational lensing \cite{sb1,RAK1}, quasinormal modes \cite{sb2}
and the accretion disk \cite{disk}.  These results could provide a
way to distinguish the deformed Ho\v{r}ava-Lifshitz theory from
standard general relativity. However, the observable properties of
black hole in the Ho\v{r}ava-Lifshitz gravity with detailed balance
condition is open in my knowledge. The main purpose of this paper is
to study the dynamical evolution of a massless scalar perturbation
in the Ho\v{r}ava-Lifshitz black-hole spacetime. We consider three
special cases in which the coupling constant $\lambda=1/3$,
$\lambda=1/2$ and $\lambda=3$ and to see whether there exists some
new features in the dynamical evolution of the perturbation.

The paper is organized as follows. In Sec.II we review in brief the
black hole solution in the Ho\v{r}ava-Lifshitz gravity. In Sec.III,
we adopt to the Horowitz-Hubeny approach to study the evolution of a
massless scalar perturbation in these background. In Sec. IV, we
present our numerical results and make a comparison between the
properties of the scalar perturbation in Ho\v{r}ava-Lifshitz black
hole and Schwarzschild AdS black hole. We summarize and discuss our
conclusions in the last section.

\section{Black holes in Ho\v{r}ava-Lifshitz gravity}

The four-dimensional metric in the ADM formalism can be expressed as
\cite{adm}
\begin{eqnarray}
 ds_{ADM}^2= - N^2 dt^2 + g_{ij} \Bigg(dx^i - N^i dt\Bigg)
\Bigg(dx^j - N^j dt\Bigg)
\end{eqnarray}
and the Ho\v{r}ava gravity action reads \cite{ho1,LMP}
\begin{eqnarray}
S_{HL}&=&\int dtd^3x \Big({\cal L}_0 + {\cal L}_1\Big),\\
{\cal L}_0 &=& \sqrt{g}N\left\{\frac{2}{\kappa^2}(K_{ij}K^{ij}
\label{action1}-\lambda K^2)+\frac{\kappa^2\mu^2(\Lambda_W R
  -3\Lambda_W^2)}{8(1-3\lambda)}\right\}\,,\\ {\cal L}_1&=&
\sqrt{g}N\left\{\frac{\kappa^2\mu^2 (1-4\lambda)}{32(1-3\lambda)}R^2
-\frac{\kappa^2}{2w^4} \left(C_{ij} -\frac{\mu w^2}{2}R_{ij}\right)
\left(C^{ij} -\frac{\mu w^2}{2}R^{ij}\right)
\right\},\label{action2}
\end{eqnarray}
where $C_{ij}$ is the Cotton tensor
\begin{eqnarray}
C^{ij}=\epsilon^{ik\ell}\nabla_k\left(R^j{}_\ell
-\frac14R\delta_\ell^j\right)=\epsilon^{ik\ell}\nabla_k R^j{}_\ell
-\frac14\epsilon^{ikj}\partial_kR\,.\label{def.K.C}
\end{eqnarray}
Taking $N^i=0$, the spherically symmetric solutions are
\cite{LMP,CCO,CLS,CY,MK,Nis,CCO1,Gho}
\begin{eqnarray}
\label{ssm} ds_{SS}^2 = - \tilde{N}^2(r)f(r)dt^2 + \frac{dr^2}{f(r)}
+ r^2 (d\theta^2 +\sin^2\theta d\varphi^2),
\end{eqnarray}
with
\begin{eqnarray}
\label{sol1} && f=1 + r^2-m\,
r^{p_\pm(\lambda)},~~~~p_\pm(\lambda)=\frac{2\lambda\pm\sqrt{6\lambda-2}}{\lambda-1},
\label{sol2} \\ && \tilde{N}=r^{q_\pm(\lambda)},~~~~q_\pm(\lambda)=
-\frac{1+3\lambda\pm2\sqrt{6\lambda-2}}{\lambda-1},
\end{eqnarray}
where $m$ is an integration constant related to the mass of the
black hole and $\lambda\ge \frac{1}{3}$. Hereafter we choose the
signs of $p(\lambda)$ and $q(\lambda)$ are negative because
$p_+(\lambda)$ and $q_+(\lambda)$ are meaningless for $\lambda=1$.
Moreover, we consider only three special cases
$\lambda=\frac{1}{3}$, $\lambda=\frac{1}{2}$ and $\lambda=3$ because
in these cases the values of $p(\lambda)$ and $q(\lambda)$ are
integers, which are convenient for us to study the dynamical
evolution of the scalar perturbations in the following calculations.
In the table I, we list the $f(r)$, $\tilde{N}(r)$ and Hawking
temperature of Ho\v{r}ava-Lifshitz black holes for
$\lambda=\frac{1}{3}$, $\lambda=\frac{1}{2}$ and $\lambda=3$
respectively.
\begin{table}[!h]
\begin{center}
\begin{tabular}{cccc}
\hline \hline\;\;\;\; $\lambda$ \;\;\;\; & \;\;\;\; $f(r) $\;\;\;\;
& \;\;\;\;  $\tilde{N}(r)$\;\;\;\;
 & \;\;\;\; $T_H$ \;\;\;\;
\\
\hline
\\
$\frac{1}{3}$& \;\;\;\;\;$1-\frac{m}{r}+r^2$\;\;\;\;\;  &
\;\;\;\; $r^3$\;\;\;\;\;
 & \;\;\;\;\;$\frac{1}{4\pi}r_+^2(3r_+^2+1)$\;\;\;\;\;
 \\
 \\
 $\frac{1}{2}$& \;\;\;\;\;$1-m+r^2$\;\;\;\;\;
  & \;\;\;\; $r$\;\;\;\;\;
 & \;\;\;\;\;$\frac{1}{2\pi}r_+^2$\;\;\;\;\;
 \\
\\
 3& \;\;\;\;\;$1-mr+r^2$\;\;\;\;\;  & \;\;\;\;$r^{-1}$\;\;\;\;\;
 & \;\;\;\;\;$\frac{1}{4\pi r_+^2}(r_+^2-1)$\;\;\;\;\;
 \\
\hline
\end{tabular}\label{tab0}
\caption{The metric functions $f(r)$, $\tilde{N}(r)$ and Hawking
temperature of Ho\v{r}ava-Lifshitz black holes for
$\lambda=\frac{1}{3}$, $\lambda=\frac{1}{2}$ and $\lambda=3$
respectively. }
\end{center}
\end{table}
For the case $\lambda=\frac{1}{3}$, although
 the expression of $f(r)$ is the same as that of Schwarzschild AdS black
 hole, its Hawking temperature is not equal to that of Schwarzschild AdS  black hole since $\tilde{N}(r)=r^3$ in the Ho\v{r}ava-Lifshitz black
 hole. This means that the thermodynamical properties of Ho\v{r}ava-Lifshitz black
 holes are completely different from those in the usual Schwarzschild AdS black
 hole. The similar discussions for $\lambda=\frac{1}{2}$ and
 $\lambda=3$ has been done in Refs. \cite{MK,Nis,CCO1,Myung}.

\section{Dynamical evolution of scalar perturbation in Ho\v{r}ava-Lifshitz black
 holes}

\begin{figure}[ht]
\begin{center}
\includegraphics[width=6cm]{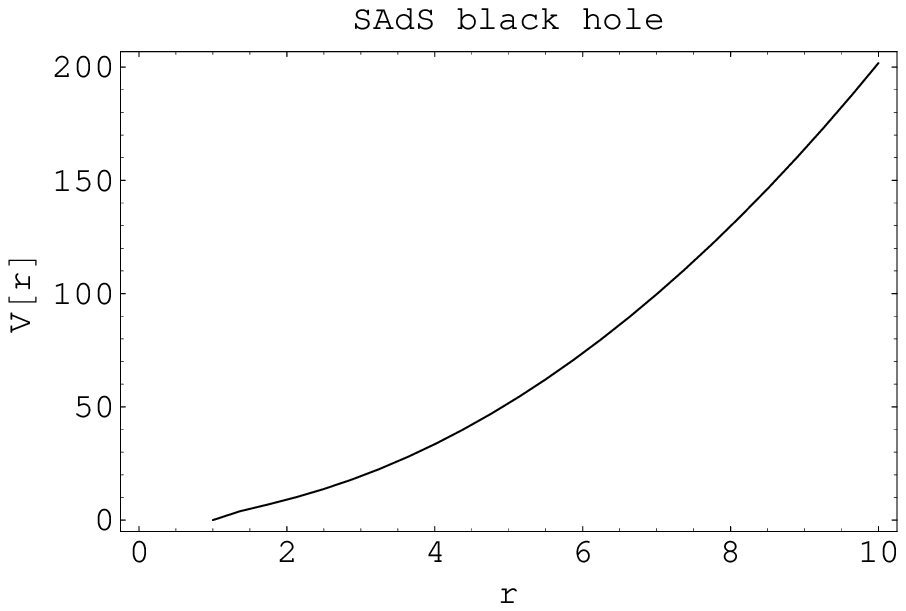}
\;\;\;\includegraphics[width=6cm]{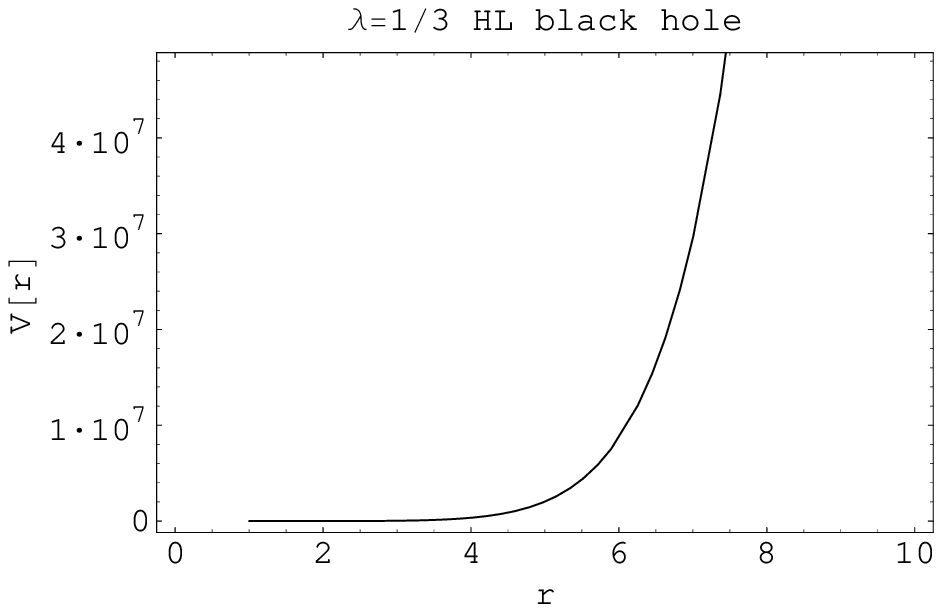}\\
 \includegraphics[width=6cm]{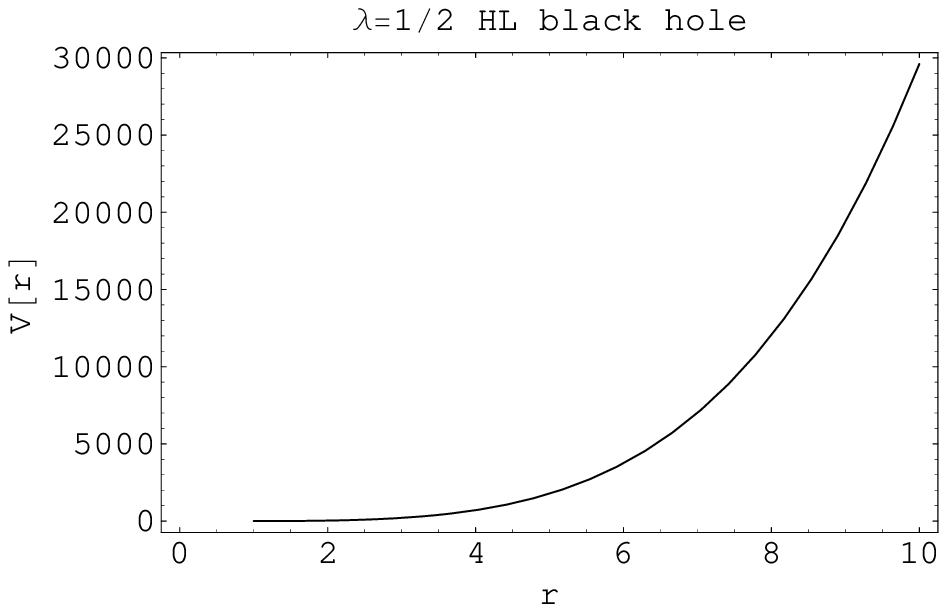}\;\;\;
\includegraphics[width=6cm]{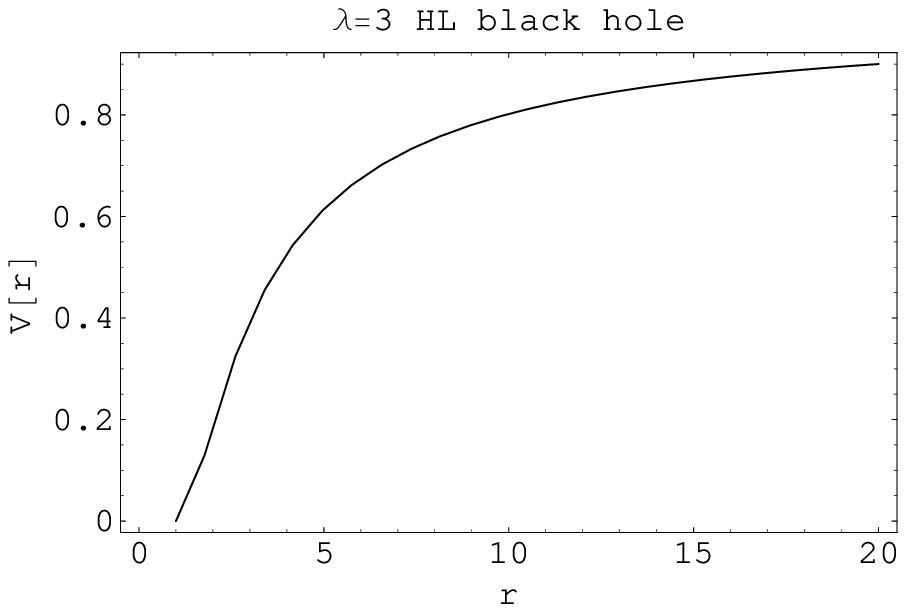}
\caption{Variety of the effective potential $V(r)$ with $r$ for
Schwarzschild AdS  black hole and the Ho\v{r}ava-Lifshitz black
holes with  $\lambda=\frac{1}{3}$, $\frac{1}{2}$ and $3$,
respectively. Here we set $l=0$ and $r_+=1$.}\label{figd}
\end{center}
\end{figure}

The evolutional equation for a massless scalar perturbation is
\begin{eqnarray}\label{k-g}
\frac{1}{\sqrt{-g}}\partial_\mu(\sqrt{-g}
g^{\mu\nu}\partial_\nu)\Phi(t,r,\theta,\varphi)=0.
\end{eqnarray}  Setting $\Phi(t,r,\theta,\varphi)=e^{-i\omega
t}R(r)Y_{lm}(\theta,\varphi)/r$ and defining the tortoise coordinate
$$dr_{*}=1/F(r)dr,$$
where $F(r)=r^{\xi /2}f(r)$ with $\xi=6,~2,-2$ for
$\lambda=\frac{1}{3}$, $\lambda=\frac{1}{2}$ and $\lambda=3$, we
find the radial equation for the scalar perturbation in the
Ho\v{r}ava-Lifshitz black-hole spacetime reads
\begin{eqnarray}\label{schrodinger}
\frac{d^2R(r)}{dr^2_{*}}+(\omega^2-V)R(r)=0,
\end{eqnarray}
with
\begin{eqnarray}
V=F(r)\bigg[\frac{1}{r}\frac{dF(r)}{dr} +\frac{l(l+1)}{r^2}r^{\xi
/2}\bigg].
\end{eqnarray}
We know from Fig. (\ref{figd}) that, for the cases
$\lambda=\frac{1}{3}$ and $\lambda=\frac{1}{2}$, the effective
potential $V(r)$ are divergent at infinity, which is similar to that
in Schwarzschild AdS black hole.  For the case $\lambda=3$, the
effective potential $V(r)$ approaches to the constant $1$.

Let us now use the Horowitz-Hubeny approach \cite{hh,wangbin} to
study the evolution of the scalar perturbations in the
Ho\v{r}ava-Lifshitz black holes above. Due to the special
asymptotical properties of their effective potentials $V(r)$ at
infinity, we take the following boundary condition
\begin{eqnarray}\label{bound}
R(r)\sim\bigg\{\begin{array}{ccc}e^{-i\omega r_\star},~~~r\rightarrow r_+,\nonumber\\
\\
0,~~~r\rightarrow +\infty,
\end{array}
\end{eqnarray}
which ensures that the solution is purely ingoing waves at the event
horizon and is not divergent at infinity at the same time.

Writing $R(r)=e^{-i\omega r_*}\phi$, Eq.  (\ref{schrodinger}) can be
rewritten as
\begin{equation}\label{three}
\frac{1}{r^{\xi /2}}\left[F\frac{d^2
\phi}{dr^2}+\big(\frac{dF}{dr}-2i\omega \big)
\frac{d\phi}{dr}-\frac{1}{r}\frac{dF}{dr}\phi\right]-\frac{l(l+1)}{r^2}\phi=0.
\end{equation}
Setting $r=\frac{1}{x}$ to change the physical region of
$r_+<r<+\infty$ into $x_+>x>0$, and then using
$B(x)=x^2F(x)=x^{2-\xi /2}f(x)$, we have
\begin{equation}\label{} x^4f(x)\frac{d^2
\phi}{dx^2}+x^2\left[(2-\frac{\xi }{2})xf(x) +x^2
\frac{df(x)}{dx}+2i \omega x^{\xi
/2}\right]\frac{d\phi}{dx}-\big[\frac{\xi
}{2}x^2f(x)-x^3\frac{df(x)}{dx} +l(l+1)x^2\big]\phi=0,
\end{equation}
which can be rewritten as
\begin{equation}\label{bx}
S(x) \frac{d^2 \phi}{dx^2}+\frac{T(x)}{x-x_+}\frac{d\phi}{dx}+
\frac{U(x)}{(x-x_+)^2}\phi=0,
\end{equation}
with
\begin{eqnarray}\label{coeff}
&&S(x)=-\frac{x^4f(x)}{(x-x_+)},\nonumber\\&&T(x)=-x^2\left[(2-\frac{\xi
}{2})x f(x) +x^2  \frac{d f(x)}{dx}+2i \omega x^{\xi
/2}\right],\nonumber\\&&
 U(x)=(x-x_+)\big[\frac{\xi }{2}x^2f(x)-x^3\frac{d f(x)}{dx}
+l(l+1)x^2\big].
\end{eqnarray}
We can expand $S(x),T(x)$ and $ U(x)$ around the point $x=x_+$ and
the corresponding coefficients of $(x-x_+)^m$ are described by
$S_m$, $T_m$ and $U_m$, respectively. Since there exists only the
ingoing modes near the event horizon, we can write $\phi$ as
\begin{equation}\label{expand}
\phi=\sum^\infty_{k=0}a_k(x-x_+)^k.
\end{equation}Substituting Eq. (\ref{expand}) into Eq. (\ref{bx}),
we can obtain a recursion relation for $a_n$:
\begin{equation}\label{coeff1}
a_n=-\frac{1}{Z_n}\sum_{k=0}^{n-1}\left[k(k-1)S_{n-k}+kT_{n-k}+U_{n-k}\right]a_k,
\end{equation}where $
Z_n=n(n-1)S_0+nT_0$. The boundary condition (\ref{bound}) at
infinity now becomes
\begin{equation}\label{an}
\sum_{k=0}^{\infty} a_k(-x_+)^k=0.
\end{equation}
In order to find the numerical solution of Eq. (\ref{an}), we have
to truncate the sum (\ref{an}) at some large $k = N$. For chosen
parameters, if the root of Eq. (\ref{an}) approaches to the same
value as $N$ increases, we can say this root is the characteristic
frequencies of the black hole. Moreover, we must increase the
precision of all the input data to avoid the ``noise" which arises
in the recursion.

\section{Numerical results}

In this section we present our numerical results for the scalar
perturbation in the Ho\v{r}ava-Lifshitz black holes with
$\lambda=\frac{1}{3}$, $\frac{1}{2}$ and $3$, respectively.

\subsection{Ho\v{r}ava-Lifshitz black hole with $\lambda=1/3$}

\begin{figure}[ht]
\begin{center}
\includegraphics[width=6cm]{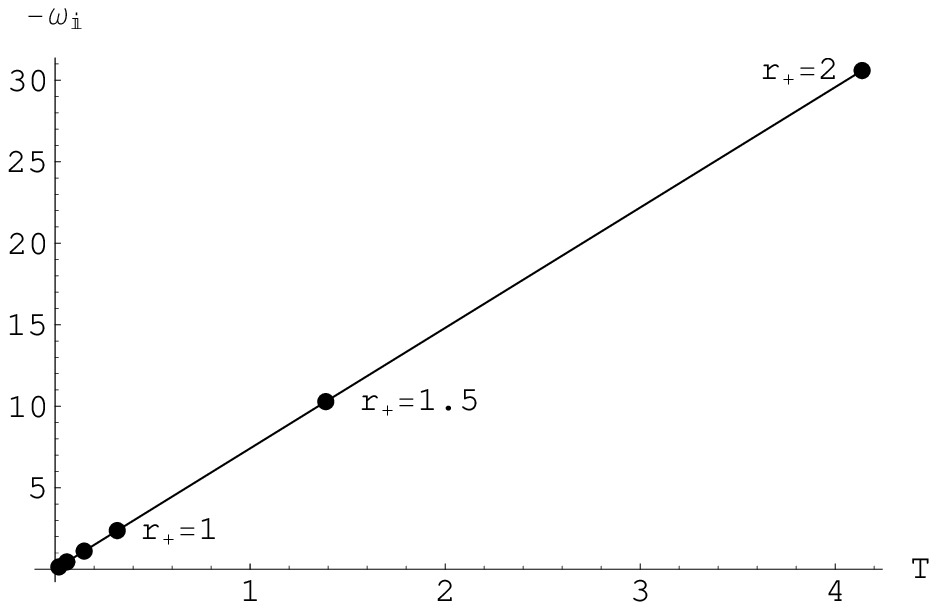}\includegraphics[width=6cm]{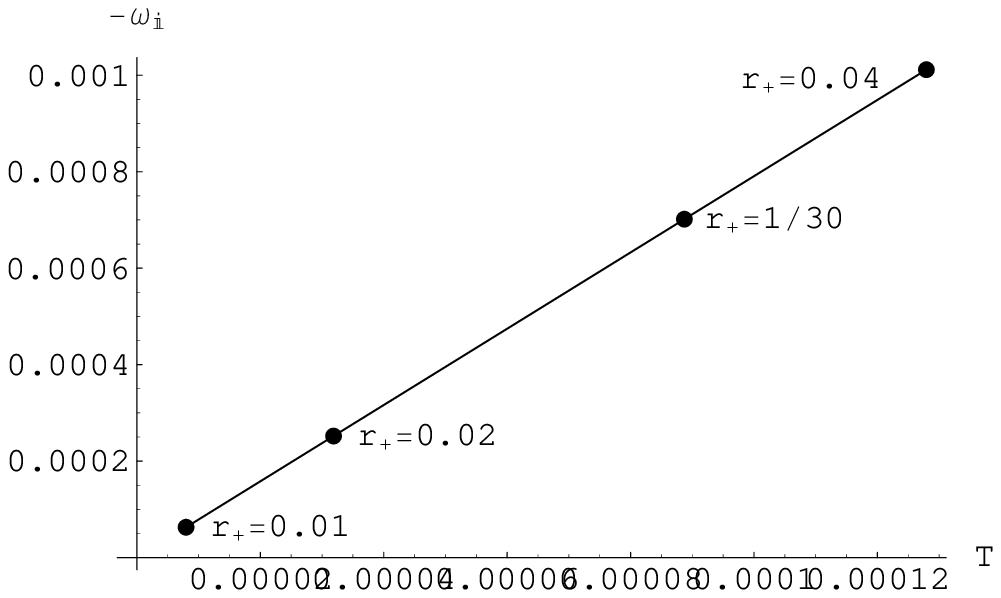}
 \caption{For Ho\v{r}ava-Lifshitz black hole with $\lambda=1/3$, $-\omega_I^{\lambda=1/3}
 $ is proportional to the temperature. The
figure is for $l=0$ case.}\label{figdd}
\end{center}
\end{figure}

In table II, we list the fundamental characteristic frequencies of
the scalar perturbation in the Ho\v{r}ava-Lifshitz black holes with
$\lambda=\frac{1}{3}$ for different values of $r_+$. From the table
II, we find that in this black hole the real parts of frequencies
vanish for the perturbations. This means that the scalar
perturbations decay without any oscillation. As the radius of the
event horizon $r_+$ increase, the absolute values of imaginary parts
increase. It implies that the scalar perturbations decay more
quickly for the larger Ho\v{r}ava-Lifshitz black hole.
\begin{table}[!h]
\begin{center}
\begin{tabular}{cccc}
\hline \hline\;\;\;\; $r_+$ \;\;\;\; & \;\;\;\; $\omega\ \ \
(l=0)$\;\;\;\;  & \;\;\;\;  $\omega \ \ \ (l=1)$\;\;\;\;
 & \;\;\;\; $\omega \ \ \ (l=2)$ \;\;\;\;
\\ \hline
100& \;\;\;\;\;-175909790i\;\;\;\;\;  & \;\;\;\;
-175911432i\;\;\;\;\;
 & \;\;\;\;\;-175914714i\;\;\;\;\;
 \\50& \;\;\;\;\;-10995509i\;\;\;\;\;
   & \;\;\;\;
-10995919i\;\;\;\;\;
 & \;\;\;\;\;-10996740i\;\;\;\;\;
 \\10& \;\;\;\;\;-17651.551i\;\;\;\;\;
  & \;\;\;\; -17667.974i\;\;\;\;\;
 & \;\;\;\;\;-17700.788i\;\;\;\;\;
  \\2& \;\;\;\;\;-30.594857i\;\;\;\;\;
  & \;\;\;\; -31.260342i\;\;\;\;\;
 & \;\;\;\;\;-32.560976i\;\;\;\;\;
 \\1.5& \;\;\;\;\;-10.284618i\;\;\;\;\;  & \;\;\;\; -10.662441i\;\;\;\;\;
 & \;\;\;\;\;-11.389131i\;\;\;\;\;
 \\
1&\;\;\;\;\;-2.3735090i\;\;\;\;\;&\;\;\;\;\;-2.5451323i\;\;\;\;\;
&\;\;\;\;\;-2.8627931i\;\;\;\;\;
 \\
0.8&\;\;\;\;\;-1.1145043i\;\;\;\;\;&\;\;\;\;\;-1.2262111i\;\;\;\;\;
&\;\;\;\;\;-1.4267695i\;\;\;\;\;
\\
0.4&\;\;\;\;\;-0.1443857i\;\;\;\;\;&\;\;\;\;\;-0.1740443i\;\;\;\;\;
&\;\;\;\;\;-0.2214635i\;\;\;\;\; \\
0.1&\;\;\;\;\;-0.0064607i\;\;\;\;\;&\;\;\;\;\;-0.0083985i\;\;\;\;\;
&\;\;\;\;\;-0.0111964i\;\;\;\;\;\\
0.04&\;\;\;\;\;-0.0010114i\;\;\;\;\;&\;\;\;\;\;-0.0013222i\;\;\;\;\;
&\;\;\;\;\;-0.0017677i\;\;\;\;\;
\\1/30&\;\;\;\;\;-0.0007015i\;\;\;\;\;&\;\;\;\;\;-0.0009173i\;\;\;\;\;
&\;\;\;\;\;-0.0012266i\;\;\;\;\;\\
0.02&\;\;\;\;\;-0.0002520i\;\;\;\;\;&\;\;\;\;\;-0.0003298i\;\;\;\;\;
&\;\;\;\;\;-0.0004410i\;\;\;\;\;
\\0.01&\;\;\;\;\;-0.0000629i\;\;\;\;\;&\;\;\;\;\;-0.0000824i\;\;\;\;\;
&\;\;\;\;\;-0.0001102i\;\;\;\;\; \\
\hline
\end{tabular}
\caption{The fundamental ($n=0$) purely damped frequencies of scalar
field in the in the $\lambda=1/3$ Ho\v{r}ava-Lifshitz black hole for
$l=0$, $1$, $2$. }
\end{center}
\end{table}
Moreover, we find that the relation between
$-\omega_I^{\lambda=1/3}$ and the temperature $T_H^{\lambda=1/3}$
can be approximated as
\begin{equation}\label{relation}
-\omega_I^{\lambda=1/3}=7.54T_H^{\lambda=1/3},
\end{equation}
which is also shown in Fig. (\ref{figdd}). This means that the
absolute value of imaginary parts is proportional to the Hawking
temperature of the black hole. It is similar to that in
4-dimensional Schwarzschild AdS black hole in which the imaginary
parts of the massless scalar perturbation can be approximated as
\cite{hh}
\begin{equation}
-\omega_I^{\text{SAdS}}=11.16T_H^{\text{SAdS}}.\label{w13}
\end{equation}
Comparing eqs. (\ref{relation}) and (\ref{w13}) we find that, if the
Hawking temperature is identical, the scalar perturbation decays in
the Ho\v{r}ava-Lifshitz black hole with $\lambda=1/3$ more slowly
than that in the Schwarzschild AdS black hole. There are other
different properties compared to Schwarzschild AdS black hole. On
one hand, the damped modes appear to increase with the angular
momentum $l$, while for Schwarzschild AdS black hole the imaginary
part of the frequencies decrease with $l$ \cite{hh}. On the other
hand, the linear relation between the imaginary part of the
frequencies and temperature is broken down for small Schwarzschild
AdS black hole, but to $\lambda=1/3$ Ho\v{r}ava-Lifshitz black hole,
this relation holds not only for large and intermediate black holes,
but also for small black hole (see Fig. \ref{figdd}).

We also study the asymptotic behavior of high overtone  of the
large, intermediate and small $\lambda=1/3$ Ho\v{r}ava-Lifshitz
black hole. In the table (III)-(V), we list the overtones of purely
damped frequencies of scalar field  for fixed $l=0,$ and $ 4$ in the
 Ho\v{r}ava-Lifshitz black hole with $\lambda=1/3$ for $r_+=100$,
 $r_+=1$ and $r_+=0.01$, respectively.
\begin{table}[!h]
\begin{center}
\begin{tabular}[b]{cccccc}
 \hline \hline
 \;\; $n$ \;\;\; &\; \;\; $\omega \  (l=0) $\;\;\;
 &  \;\;\;$\omega \  (l=4) $ \;\;\; &\;\;\;
  $n$\;\;\;
 & \; \;\;$\omega \  (l=0)$ \;\;\;& \;\;\; $\omega  \ (l=4) $\;\;\; \\ \hline
\\
0& -175909790i& -175926204i&  8& -1384766991i& -1384773476i
 \\
1& -329818196i &-329830731i&  9 & -1534973359i& -1534979534i
 \\
2&-481590494i&-481601084i&10&-1685148584i&-1685154489i
 \\
3&-632632175i&-632641529i&11&-1835299632i&-1835305301i
 \\
4&-783328595i&-783337071i&12&-1985431502i&-1985436962i
\\
5&-933832036i&-933839846i&13&-2135547882i&-2135553153i
\\6&-1084215699i&-1084222980i&14&-2285651553i&-2285656654i
\\7&-1234519450i&-1234526298i&15&-2435744657i&-2435749604i
\\
\hline
\end{tabular}
\label{t23} \caption{The overtones of purely damped frequencies of
scalar field in the Ho\v{r}ava-Lifshitz black hole with
$\lambda=1/3$ for $r_+=100,l=0,4$. The asymptotic behavior is
$\omega^{\lambda=1/3}=-(1.506n+1.767)r_+^4i$ for any value of $l$.}
\end{center}
\end{table}
\begin{table}[!h]
\begin{center}
\begin{tabular}[b]{cccccc}
 \hline \hline
 \;\;\; $n$ \;\;\; &\;\;\; $\omega \ (l=0)  $\;\;\; &\;\;\; $\omega\ (l=4)  $\;\;\;
  & \;\;\;  $n$\;\;\;
 & \;\;\; $\omega \ (l=0)$ \;\;\; & \;\;\; $\omega \ (l=4) $ \;\;\;\\ \hline
\\0& -2.373509i   & -3.778299i &9& -20.491285i& -21.171191i
 \\
1& -4.428416i  & -5.624912i& 10 & -22.492943i& -23.145882i
 \\
2&-6.451969i& -7.515653i&11&-24.494323i& -25.123273i
 \\
3&-8.465115i& -9.433003i&12&-26.495486i& -27.102915i
 \\
4&-10.473494i& -11.367757i&13&-28.496477i& -29.084457i
\\
5&-12.479287i& -13.314594i&14&-30.497331i& -31.067620i
\\6&-14.483518i& -15.270211i&15&-32.498073i& -33.052181i
\\7&-16.486736i& -17.232437i&16&-34.498723i& -35.037955i
\\8&-18.489259i& -19.199784i&17&-36.499293i& -37.024791i
\\
\hline
\end{tabular}\label{t21}
\caption{The overtones of purely damped frequencies of scalar field
in the Ho\v{r}ava-Lifshitz black hole with $\lambda=1/3$ for
$r_+=1,~l=0,~4$. For large $n$ one finds that
 $\omega^{\lambda=1/3}\sim-(2.007n+2.906)i$ for $l=0$.}
\end{center}
\end{table}
\begin{table}[!h]
\begin{center}
\begin{tabular}[b]{cccccc}
 \hline \hline
 \;\;\; $n$ \;\;\; & \;\;\; $\omega \ (l=0) $\;\;\;& \;\;\; $\omega \ (l=4) $\;\;\;
   & \;\;\;  $n$\;\;\;
 & \;\;\; $\omega \ (l=0)$ \;\;\;& \;\;\; $\omega \ (l=4) $\;\;\; \\ \hline
\\0&  -0.0000629i &  -0.0001721i  &   9&  -0.0005206i&  -0.0006121i
 \\
1&  -0.0001158i &  -0.0002199i  &   10 &  -0.0005708i&  -0.0006616i
 \\
2&-0.0001672i&  -0.0002682i&11&-0.0006210i&  -0.0007111i
 \\
3&-0.0002182i&  -0.0003168i&12&-0.0006712i&  -0.0007607i
 \\
4&-0.0002688i&  -0.0003657i&13&-0.0007213i&  -0.0008104i
\\
5&-0.0003193i&  -0.0004147i&14&-0.0007715i&  -0.0008600i
\\6&-0.0003698i&  -0.0004639i&15&-0.0008216i&  -0.0009097i
\\7&-0.0004201i&  -0.0005132i&16&-0.0008717i&  -0.0009594i
\\8&-0.0004704i&  -0.0005626i&17&-0.0009218i&  -0.0010092i
\\
\hline
\end{tabular}\label{t22}
\caption{The overtones of purely damped frequencies of scalar field
in the Ho\v{r}ava-Lifshitz black hole with $\lambda=1/3$ for
$r_+=0.01,~l=0,~4$. For large $n$ one finds that
$\omega^{\lambda=1/3}\sim-(5.05n+6.33)\times10^{-5}i$ for $l=0$.}
\end{center}
\end{table}
As the overtone number $n$ increases, the absolute value of
imaginary parts increase and the modes are evenly spaced, which is
similar to that in the usual black-hole spacetimes. In the large
Ho\v{r}ava-Lifshitz black hole with $\lambda=1/3$ (see table III),
the evenly spaced frequencies of the perturbations is approximated
as
\begin{equation}\omega^{\lambda=1/3}=-\frac{4\pi}{3}(1.506n+1.767)T_H^{\lambda=1/3}i. \end{equation} This asymptotic behavior for the spacing holds for
any value of $l$, i.e. this spacing is $l$-independent as the same
as Schwarzschild AdS black hole \cite{cardoso}.

\subsection{Ho\v{r}ava-Lifshitz black hole with $\lambda=\frac{1}{2}$ }

\begin{figure}[ht]
\begin{center}
\includegraphics[width=6cm]{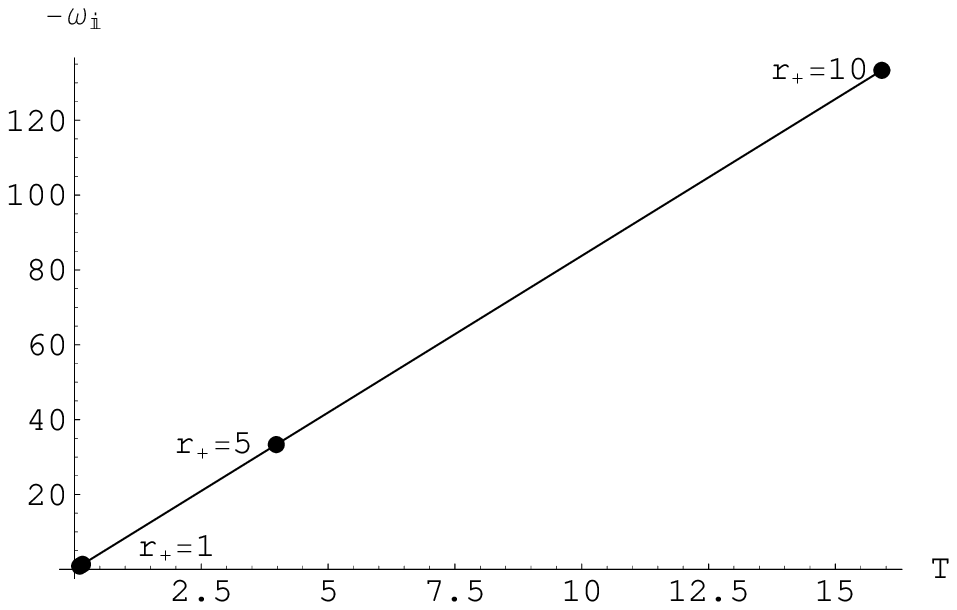}\;\includegraphics[width=6cm]{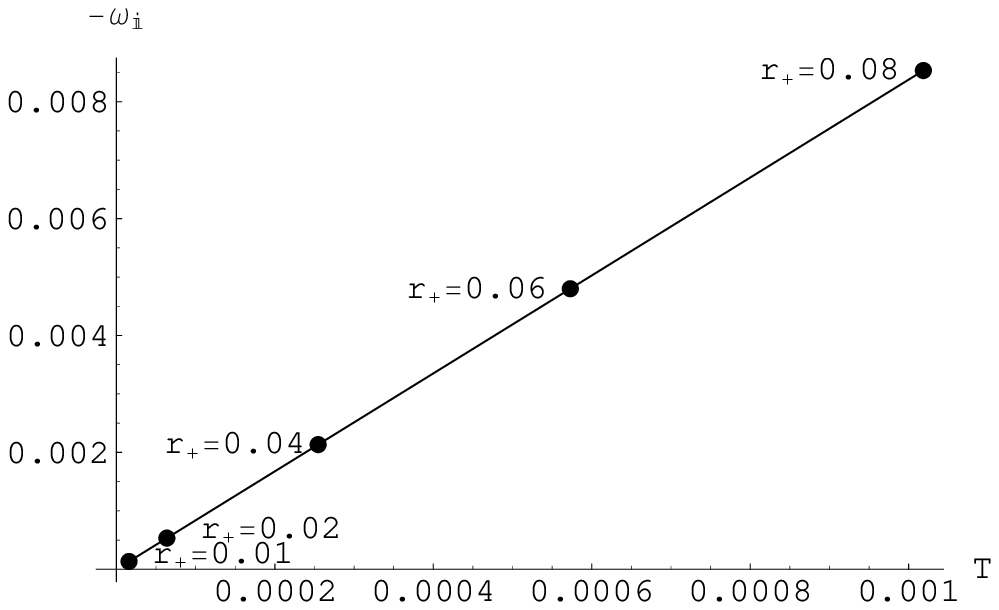}
 \caption{For $\lambda=1/2$ Ho\v{r}ava-Lifshitz black hole,
  $-\omega_I$ is proportional to the Hawking temperature. The
figure is for $l=0$ case.}\label{fig05}
\end{center}
\end{figure}

Now let us consider the scalar perturbation in the
Ho\v{r}ava-Lifshitz black hole with $\lambda=1/2$. In table VI, we
list the fundamental characteristic frequencies of the scalar
perturbation for different values of $r_+$. Similarly, we find  that
in this black hole the real parts of frequencies also disappear for
the perturbations (as shown in Fig. (\ref{fig05})). Moreover, we
also find the absolute values of imaginary parts increase with the
radius of the event horizon $r_+$. It implies that in this case the
scalar perturbations decay more quickly in the larger black hole.
\begin{table}[!h]
\begin{center}
\begin{tabular}[b]{cccc}
 \hline \hline
 \;\;\;\; $r_+$ \;\;\;\; & \;\;\;\; $\omega\ \ \ (l=0)$\;\;\;\;  & \;\;\;\;  $\omega \ \ \ (l=1)$\;\;\;\;
 & \;\;\;\; $\omega \ \ \ (l=2)$ \;\;\;\; \\ \hline
 \\100& \;\;\;\;\;-13333.333i\;\;\;\;\;  & \;\;\;\;
-13333.666i\;\;\;\;\;
 & \;\;\;\;\;-13334.333i\;\;\;\;\;
 \\
50& \;\;\;\;\;-3333.3333i\;\;\;\;\;  & \;\;\;\;
-3333.6666i\;\;\;\;\;
 & \;\;\;\;\;-3334.3333i\;\;\;\;\;
\\10& \;\;\;\;\;-133.33333i\;\;\;\;\;  & \;\;\;\; -133.66666i\;\;\;\;\;
 & \;\;\;\;\;-134.33333i\;\;\;\;\;
 \\
5& \;\;\;\;\;-33.333333i\;\;\;\;\;  & \;\;\;\; -33.666666i\;\;\;\;\;
 & \;\;\;\;\;-34.333333i\;\;\;\;\;
 \\
1&-1.333333i&-1.666666i&-2.333333i
 \\
0.8&-0.853333i&-1.186666i&-1.853333i
 \\
0.6&-0.480000i&-0.813333i&-1.464000i
\\
0.4&-0.213333i&-0.546666i&-0.977142i
\\
\hline
\end{tabular}\label{tab2}
\caption{The fundamental ($n=0$) purely damped frequencies of scalar
field in the in the $\lambda=\frac{1}{2}$ Ho\v{r}ava-Lifshitz black
hole for fixed $l=0$, $1$, $2$.}
\end{center}
\end{table}
For the Ho\v{r}ava-Lifshitz black hole with $\lambda=\frac{1}{2}$,
the relation between $-\omega_I^{\lambda=1/2}$ and the Hawking
temperature reads approximately
\begin{equation}\label{relationn}
-\omega_I^{\lambda=1/2}=8.378T_H^{\lambda=1/2}.
\end{equation}
This relation holds not only for large and intermediate black holes,
but also for small black hole.  Comparing it with Eq.(\ref{w13}), we
find, as the Hawking temperature is identical, the scalar
perturbation decays more slowly in the Ho\v{r}ava-Lifshitz black
hole with $\lambda=1/2$.
Similarly, in the table (VII)-(IX), we list the overtones of purely
damped frequencies of scalar field  for fixed $l=0$ and $l=4$ in the
 Ho\v{r}ava-Lifshitz black hole with $\lambda=1/2$ for $r_+=100$,
 $r_+=1$ and $r_+=0.4$, respectively.
 We find that the modes
are evenly spaced for large black hole (see table VII)
\begin{equation}\omega^{\lambda=1/2}=-2\pi(1.009n+1.333)T_H^{\lambda=1/2}i.\end{equation}  This asymptotic behavior for the spacing
holds for any value of $l$, i.e. this spacing is also
$l$-independent as the same as $\lambda=1/3$ Ho\v{r}ava-Lifshitz
black hole.
\begin{table}[!h]
\begin{center}
\begin{tabular}[b]{cccccc}
 \hline \hline
 \;\;\; $n$ \;\;\; & \;\;\; $\omega\ (l=0) $\;\;\;& \;\;\; $\omega\ (l=4) $
 \;\;\;  &\;\;\;  $n$\;\;\;
 & \;\;\; $\omega\ (l=0) $ \;\;\;& \;\;\; $\omega\ (l=4) $\;\;\; \\ \hline
\\0& -13333.333i & -13336.666i& 9 & -104761.90i& -104762.38i
 \\
1& -24000.000i & -24002.000i & 10 & -114782.60i& -114783.04i
 \\
2&-34285.714i& -34287.142i&11&-124800.00i& -124800.40i
 \\
3&-44444.444i& -44445.555i&12&-134814.81i& -134815.18i
 \\
4&-54545.454i& -54546.363i&13&-144827.58i& -144827.93i
\\
5&-64615.384i& -64616.153i&14&-154838.70i& -154839.03i
\\6&-74666.666i& -74667.333i&15&-164848.48i& -164848.78i
\\7&-84705.882i& -84706.470i&16&-174857.14i& -174857.42i
\\8&-94736.842i& -94737.368i&17&-184864.86i& -184865.13i
\\
\hline
\end{tabular}\label{tab31}
\caption{The overtones of purely damped frequencies of scalar field
in the Ho\v{r}ava-Lifshitz black hole with $\lambda=1/2$ for
$r_+=100,l=0,4.$ For large black hole one finds approximately
$\omega^{\lambda=1/2}=-(1.009n+1.333)r_+^2i$ for any value of $l$. }
\end{center}
\end{table}
\begin{table}[!h]
\begin{center}
\begin{tabular}[b]{cccccc}
 \hline \hline
 \;\;\; $n$ \;\;\; & \;\;\; $\omega\ (l=0) $\;\;\;& \;\;\; $\omega\ (l=4) $
 \;\;\;  &\;\;\;  $n$\;\;\;
 & \;\;\; $\omega\ (l=0) $ \;\;\;& \;\;\; $\omega\ (l=4) $\;\;\;  \\ \hline
\\0& -1.333333i & -4.400000i &9 & -10.47619i& -10.95238i
 \\
1& -2.400000i & -4.666666i &  10 & -11.47826i& -11.91304i
 \\
2&-3.428571i& -4.857142i&11&-12.48000i& -12.88000i
 \\
3&-4.444444i& -5.555555i&12&-13.48148i& -13.85185i
 \\
4&-5.454545i& -6.363636i&13&-14.48275i& -14.82758i
\\
5&-6.461538i& -7.230769i&14&-15.48387i& -15.80645i
\\6&-7.466666i& -8.133333i&15&-16.48484i& -16.78787i
\\7&-8.470588i& -9.058823i&16&-17.48571i& -17.77142i
\\8&-9.473684i& -10.00000i&17&-18.48648i& -18.75675i
\\
\hline
\end{tabular}\label{tab32}
\caption{The overtones of purely damped frequencies of scalar field
in the Ho\v{r}ava-Lifshitz black hole with $\lambda=1/2$ for
$r_+=1,l=0,~4$. Asymptotically for large $n$ one finds approximately
$\omega^{\lambda=1/2}\sim-(1.009n+1.333)i$ for $l=0$.}
\end{center}
\end{table}
\begin{table}[!h]
\begin{center}
\begin{tabular}[b]{cccccc}
 \hline \hline
 \;\;\; $n$ \;\;\; & \;\;\; $\omega\ (l=0) $\;\;\;& \;\;\; $\omega\ (l=4) $
 \;\;\;  &\;\;\;  $n$\;\;\;
 & \;\;\; $\omega\ (l=0) $ \;\;\;& \;\;\; $\omega\ (l=4) $\;\;\; \\ \hline
\\0& -0.213333i  & -1.781818i  &  9& -1.676190i& -2.384000i
 \\
1&-0.384000i& -1.803076i   &  10 & -1.836521i& -2.396800i
 \\
2&-0.548571i& -1.82222i  &11&-1.996800i& -2.527407i
 \\
3&-0.711111i& -1.861333i  &12&-2.157037i& -2.662068i
 \\
4&-0.872727i& -1.943529i  &13&-2.317241i& -2.800000i
\\
5&-1.033846i& -1.977142i  &14&-2.477419i& -2.940606i
\\6&-1.194666i& -2.042105i  &15&-2.637575i& -3.083428i
\\7&-1.355294i& -2.152380i  &16&-2.797714i& -3.228108i
\\8&-1.515789i& -2.271304i  &17&-2.957837i& -3.374358i
\\
\hline
\end{tabular}\label{tab33}
\caption{The overtones of purely damped frequencies of scalar field
in the Ho\v{r}ava-Lifshitz black hole with $\lambda=1/2$  for
$r_+=0.4,l=0,~4$. Asymptotically for large $n$ one finds
approximately $\omega^{\lambda=1/2}\sim-(0.161n+0.221)i$ for $l=0$.}
\end{center}
\end{table}

\subsection{Ho\v{r}ava-Lifshitz black hole with $\lambda=3$}

For $\lambda=3$, there exists extremal case with $r_+=1$ because the
temperature vanishes. It is thermodynamically unstable even for
large black hole \cite{CCO1} and shows some marginal properties. For
example, the small black holes ($r_+<1$) have no physical meaning
since the temperature is negative, while for large black holes the
temperature approach to a constant
$T_H^{\lambda=3}\rightarrow\frac{1}{4\pi}$. Therefore the decay of
scalar perturbation will different from previous cases $\lambda=1/3$
and $\lambda=1/2$.

By repeating the previous operation, we find that it is  very
difficult to find the convergent root.  Under the condition of
$100$-digital precision of input data, there is no convergent root.
So we improve precision to $300$-digital precision, the convergent
roots appear but the convergence becomes much slower than
$\lambda=1/3$ and $\lambda=1/2$ cases. It is also difficult to seek
high overtone.
\begin{table}[!h]
\begin{center}
\begin{tabular}[b]{cccc}
 \hline \hline
 \;\;\;\; $r_+$ \;\;\;\; & \;\;\;\;\;\;
  $\omega\ \ \ (l=0)$\;\;\;\;
   & \;\;\;\; $r_+$ \;\;\;\;& \;\;\;\; $\omega\ \ \ (l=0)$\;\;\;\;
    \\ \hline
\\
100&-70.252656i&4&-2.353367i
 \\
 50&   -34.897542i & \;3&-1.630615i  \\
10& -6.611997i& 2& -0.880420i \\
5& -3.067646i  \\
\hline
\end{tabular}\label{tab3}
\caption{The fundamental ($n=0$) damped frequencies of scalar field
in the Ho\v{r}ava-Lifshitz black hole with $\lambda=3$ for $l=0$.}
\end{center}
\end{table}
\begin{figure}[ht]
\begin{center}
\includegraphics[width=6cm]{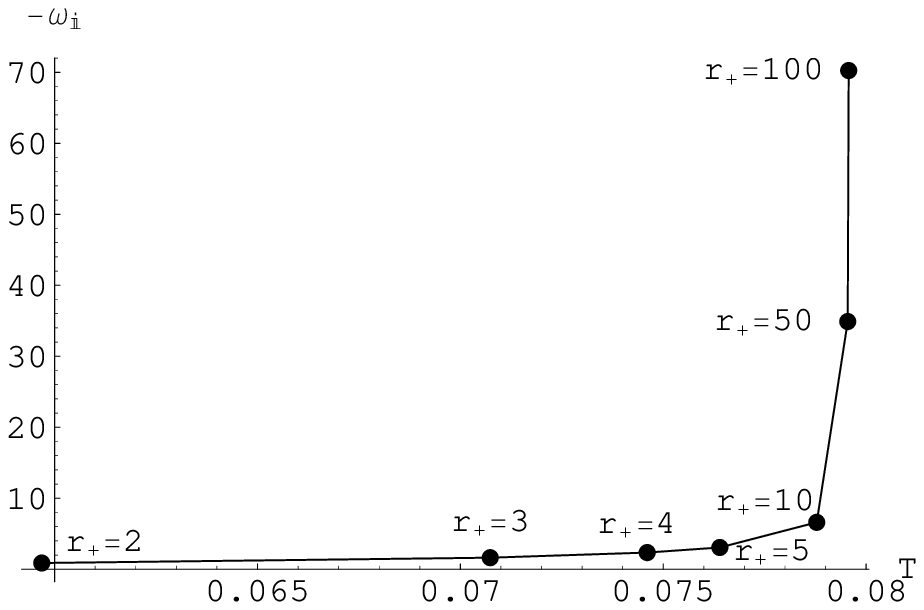}\;\includegraphics[width=6cm]{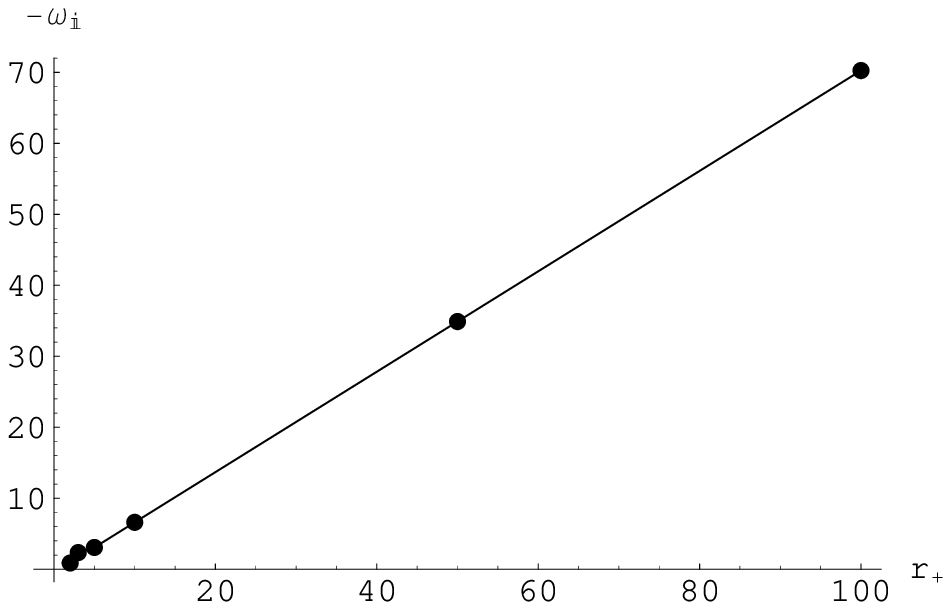}
\;\;\;\
 \caption{ The purely damped modes for
the scalar perturbations in the Ho\v{r}ava-Lifshitz black hole with
$\lambda=3$ and $l=0$. Though the modes do not linear to the
temperature (left), they linear to the radii of events horizon,
i.e., $ -\omega_I^{\lambda=3}=0.669r_+$ (right).}\label{fig3}
\end{center}
\end{figure}

In the table (X) we list the characteristic frequencies of the
scalar perturbation in the Ho\v{r}ava-Lifshitz black holes with
$\lambda=3$ for different values of $r_+$. From the table (X), it is
easy to see that the modes are not linear to the temperature at all,
but they are still monotonically increasing with the temperature
(see Fig. (\ref{fig3})).  Though the modes do not linear to the
temperature, they linear to the radii of horizon in large black hole
region \begin{eqnarray} -\omega_I^{\lambda=3}=0.669r_+ .
\end{eqnarray} This property is similar to Schwarzschild AdS black
hole, $-\omega_I^{SAdS}=2.66r_+$ \cite{hh}.

\section{Conclusions and discussions }

We study the dynamical evolution of the massless scalar perturbation
in the background of Ho\v{r}ava-Lifshitz black-hole spacetime. For
the cases of dynamical coupling constants $\lambda=1/3$, $1/2$ and
$3$, there are purely damped modes which are different from those in
the usual black-hole spacetimes. For $\lambda=1/3$ and $1/2$ cases
the imaginary parts of frequency are proportional to the Hawking
temperature not only for large and intermediate black holes, but
also for small black hole. If the black holes possess the same
Hawking temperatures, the scalar perturbation decays most quickly in
the Schwarzschild AdS black-hole spacetime. For $\lambda=3$ case,
the imaginary parts of frequency are not proportional to the Hawking
temperature any more, but is linear to the radii of event horizon of
the black hole.

Why there are only purely damped modes in three kinds of
Ho\v{r}ava-Lifshitz black holes? Let us recall that purely damped
mode is also appeared in Einstein's gravity, for example, some modes
of the Schwarzschild AdS black hole due to Dirac \cite{jingm},
electromagnetic and gravitational \cite{cardoso1,cardoso}
perturbations and Reissner-Nordstr\"{o}m AdS black hole due to
electromagnetic, gravitational \cite{berti} and Rarita-Schwinger
\cite{zhangyun} field perturbations for low overtone numbers. But
for scalar field, the purely damped mode has never appeared in
Einstein's gravity. In classical mechanics, the over-damped
condition will lead to a purely damped mode. By comparing the wave
equation (\ref{schrodinger}) to damped wave equation in classical
mechanics, we find that the potential $V(r)$ is exact damped term
which results in energy dissipation. From the Fig. (\ref{figd}), one
can see that the potential of scalar perturbation for $\lambda=1/3$
Ho\v{r}ava-Lifshitz black hole is proportional to $r^8$ at long
distance, for $\lambda=1/2$ Ho\v{r}ava-Lifshitz black hole it is
$r^4$, but for Schwarzschild AdS black hole is just $r^2$.  $r^4$
and $r^8$ asymptotic potentials may be over-damped potentials, so
that the decay with no oscillation for $\lambda=1/3$ and
$\lambda=1/2$ Ho\v{r}ava-Lifshitz black holes.

\begin{figure}[ht]
\begin{center}
\includegraphics[width=6cm]{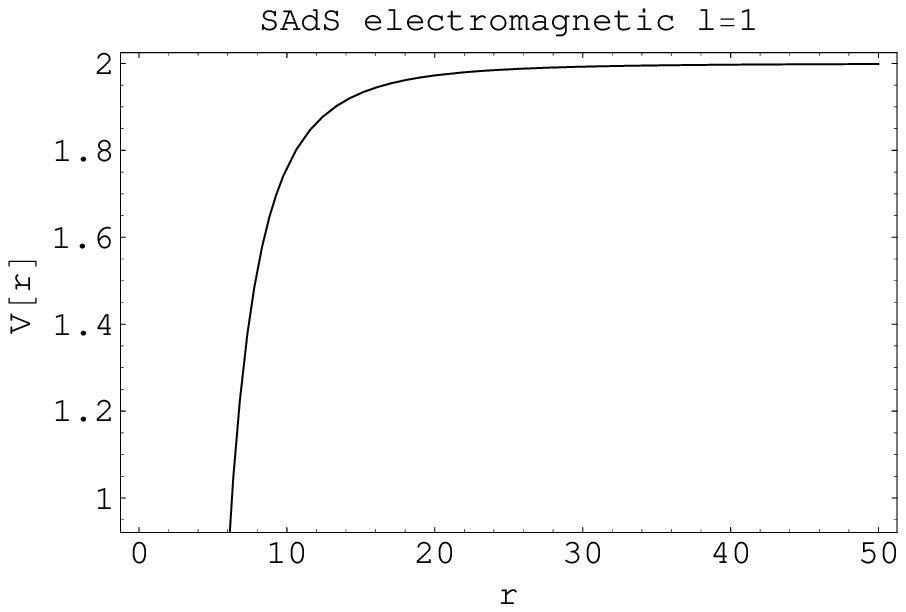}
\includegraphics[width=6cm]{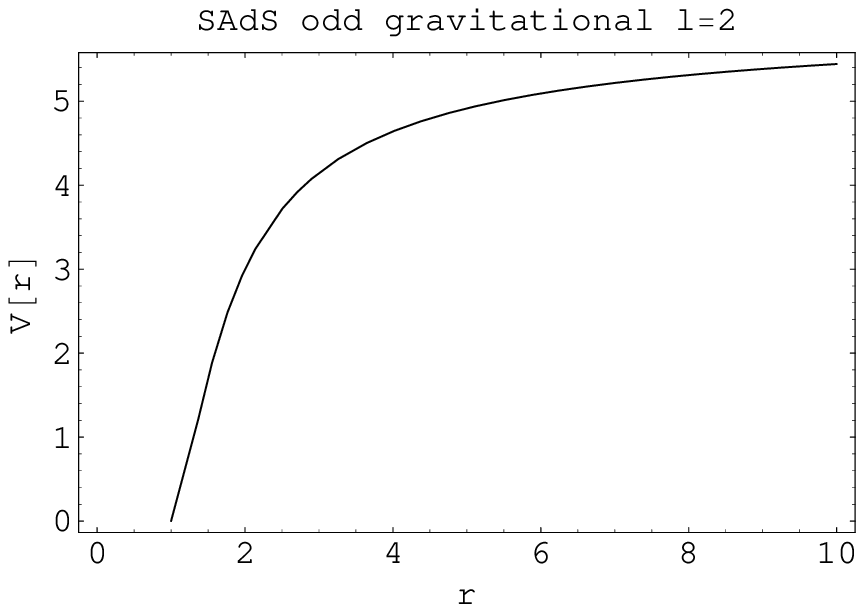}\;\;\;
 \caption{ The potential for
 the electromagnetic perturbation with $l=1,~r_+=5$ (left), and for
the odd gravitational perturbation with $l=2,~r_+=1$ (right) in the
Schwarzschild AdS black hole.}\label{fig4}
\end{center}
\end{figure}

From the Fig. (\ref{fig4}) we know that the potential of
electromagnetic perturbation with $l=1$ in Schwarzschild AdS black
hole is convergent to constant $2$ at infinity and potential of odd
gravitational perturbation with $l=2$ is $6$
\cite{cardoso1,cardoso}, while for $\lambda=3$ Ho\v{r}ava-Lifshitz
black hole it converges to constant $1$ at infinity (Fig.
(\ref{figd})). The authors in Ref. \cite{cardoso1} find that, due to
electromagnetic perturbation, there exist the lowest purely damped
modes so long as $r_+\geq5$ with angular momentum $l=1$, and
$r_+\geq10$ with $l=2$; to odd gravitational perturbation, there
exist the lowest purely damped modes if only $r_+\geq0.5$ with
$l=2$, and $r_+\geq1$ with $l=3$. In ref. \cite{cardoso}, the
authors find that there exist eight purely damped modes for
$r_+=1000$, four for $r_+=100$, and two for $r_+=10$. Then they
conclude that ``infinitely large black holes may have pure imaginary
electromagnetic QN frequencies for any overtone number". Therefore,
it is understandable that there exist purely damped modes in
$\lambda=3$ Ho\v{r}ava-Lifshitz black hole.

\begin{acknowledgments}
This work was supported by the National Natural Science Foundation
of China under Grant No 10875040; the FANEDD under Grant No. 200317,
the Hunan Provincial Natural Science Foundation of China under Grant
No. 08JJ0001, the Hunan Provincial Innovation Foundation for
Postgraduate,  and Construct Program of the National Key Discipline.
\end{acknowledgments}

\end{document}